\documentclass[aps,prl,preprintnumbers,amsmath,amssymb,latexsym,nofootinbib,array,enumerate,letter,twocolumn,superscriptaddress]{revtex4}

\usepackage{graphicx}
\usepackage{epstopdf}
\usepackage{dcolumn}
\usepackage{bm}
\usepackage[colorlinks=false]{hyperref}
\usepackage[dvipsnames]{xcolor}
\usepackage{amsmath}

\begin{document}


\def\a{\alpha}
\def\b{\beta}
\def\c{\varepsilon}
\def\d{\delta}
\def\e{\epsilon}
\def\f{\phi}
\def\g{\gamma}
\def\h{\theta}
\def\k{\kappa}
\def\l{\lambda}
\def\m{\mu}
\def\n{\nu}
\def\p{\psi}
\def\q{\partial}
\def\r{\rho}
\def\s{\sigma}
\def\t{\tau}
\def\u{\upsilon}
\def\v{\varphi}
\def\w{\omega}
\def\x{\xi}
\def\y{\eta}
\def\z{\zeta}
\def\D{\Delta}
\def\G{\Gamma}
\def\H{\Theta}
\def\L{\Lambda}
\def\F{\Phi}
\def\P{\Psi}
\def\S{\Sigma}

\def\o{\over}
\def\beq{\begin{align}}
\def\eeq{\end{align}}
\newcommand{\gsim}{ \mathop{}_{\textstyle \sim}^{\textstyle >} }
\newcommand{\lsim}{ \mathop{}_{\textstyle \sim}^{\textstyle <} }
\newcommand{\vev}[1]{ \left\langle {#1} \right\rangle }
\newcommand{\bra}[1]{ \langle {#1} | }
\newcommand{\ket}[1]{ | {#1} \rangle }
\newcommand{\mEV}{ {\rm meV} }
\newcommand{\EV}{ {\rm eV} }
\newcommand{\KEV}{ {\rm keV} }
\newcommand{\MEV}{ {\rm MeV} }
\newcommand{\GEV}{ {\rm GeV} }
\newcommand{\TEV}{ {\rm TeV} }
\newcommand{\1}{\mbox{1}\hspace{-0.25em}\mbox{l}}
\newcommand{\headline}[1]{\noindent{\bf #1}}
\def\diag{\mathop{\rm diag}\nolimits}
\def\Spin{\mathop{\rm Spin}}
\def\SO{\mathop{\rm SO}}
\def\O{\mathop{\rm O}}
\def\SU{\mathop{\rm SU}}
\def\U{\mathop{\rm U}}
\def\Sp{\mathop{\rm Sp}}
\def\SL{\mathop{\rm SL}}
\def\tr{\mathop{\rm tr}}
\def\mpl{M_{\rm Pl}}

\def\IJMP{Int.~J.~Mod.~Phys. }
\def\MPL{Mod.~Phys.~Lett. }
\def\NP{Nucl.~Phys. }
\def\PL{Phys.~Lett. }
\def\PR{Phys.~Rev. }
\def\PRL{Phys.~Rev.~Lett. }
\def\PTP{Prog.~Theor.~Phys. }
\def\ZP{Z.~Phys. }

\def\dd{\mathrm{d}}
\def\ff{\mathrm{f}}
\def\BH{{\rm BH}}
\def\inf{{\rm inf}}
\def\ev{{\rm evap}}
\def\eq{{\rm eq}}
\def\SM{{\rm sm}}
\def\Mpl{M_{\rm Pl}}
\def\GeV{{\rm GeV}}

\def\newpar{\vskip4pt}

\title{
Dark Photon Dark Matter Produced by Axion Oscillations
}
\preprint{LCTP-18-21}

\author{Raymond T.~Co}
\affiliation{Leinweber Center for Theoretical Physics, Department of Physics, University of Michigan, Ann Arbor, MI 48109, USA}
\author{Aaron Pierce}
\affiliation{Leinweber Center for Theoretical Physics, Department of Physics, University of Michigan, Ann Arbor, MI 48109, USA}
\author{Zhengkang Zhang}
\affiliation{Leinweber Center for Theoretical Physics, Department of Physics, University of Michigan, Ann Arbor, MI 48109, USA}
\affiliation{Department of Physics, University of California, Berkeley, CA 94720, USA}
\affiliation{Theoretical Physics Group, Lawrence Berkeley National Laboratory, Berkeley, CA 94720, USA}

\author{Yue Zhao}
\affiliation{Leinweber Center for Theoretical Physics, Department of Physics, University of Michigan, Ann Arbor, MI 48109, USA}
\affiliation{Department of Physics and Astronomy, University of Utah, Salt Lake City, UT 84112, USA}

\begin{abstract}
Despite growing interest and extensive effort to search for ultralight dark matter in the form of a hypothetical dark photon, how it fits into a consistent cosmology is unclear. Several dark photon dark matter production mechanisms proposed previously are known to have limitations, at least in certain mass regimes of experimental interest. In this letter, we explore a novel mechanism, where a coherently oscillating axion-like field can efficiently transfer its energy density to a dark photon field via a tachyonic instability. The residual axion relic is subsequently depleted via couplings to the visible sector, leaving only the dark photon as dark matter. We ensure that the cosmologies of both the axion and dark photon are consistent with existing constraints.  We find that the mechanism works for a broad range of dark photon masses, including those of interest for ongoing experiments and proposed detection techniques.
\end{abstract}

\date{\today}

\maketitle

{\bf Introduction.}---%
%
The identity of dark matter remains unknown.  One candidate is a dark photon $A'$, a novel gauge boson with an unknown mass and tiny coupling to the Standard Model (SM).  Recent years have seen a growing interest in experimental techniques for detecting light dark photon dark matter (DPDM), and many ideas have been studied. These include microwave cavity experiments such as ADMX \cite{Wagner:2010mi}, a dark matter radio \cite{Chaudhuri:2014dla}, dish antennas \cite{ Horns:2012jf,Jaeckel:2013sqa,Suzuki:2015sza,Jaeckel:2015kea,Dobrich:2015tpa,Knirck:2018ojz}, dielectric haloscopes \cite{Baryakhtar:2018doz}, absorption in various targets \cite{Hochberg:2016ajh, Hochberg:2016sqx,Yang:2016zaz, Bloch:2016sjj, Bunting:2017net, Hochberg:2017wce, Arvanitaki:2017nhi, Knapen:2017ekk, Griffin:2018bjn}, the use of dark matter detectors as helioscopes \cite{An:2013yua},  and repurposing of  gravitational wave detectors \cite{Pierce:2018xmy} as well as other accelerometers \cite{Graham:2015ifn}. All these techniques focus on DPDM with masses  $\lesssim$ keV, in which case it must be produced nonthermally to avoid constraints on warm dark matter \cite{Irsic:2017ixq,Lopez-Honorez:2017csg}. It is thus important to understand precisely how this could occur.

Several mechanisms for nonthermal production of DPDM have been studied in the literature. One possibility is a misalignment mechanism \cite{Nelson:2011sf} similar to that of axions: $A'$ is initially displaced from the minimum of its potential, and the energy stored in its oscillations may act as dark matter. It was later pointed out, however, that sufficient production of DPDM via this mechanism is difficult absent  a nonminimal coupling of $A'$ to gravity \cite{Arias:2012az}, which introduces a quadratic divergence for $m_{A'}$ \cite{Graham:2015rva}. An alternative approach relying on quantum fluctuations during inflation can realize DPDM with $m_{A'}>10^{-5}\,$eV \cite{Graham:2015rva}, but the St\"{u}ckelberg mass and high scale inflation essential to the mechanism may be in tension with the Weak Gravity Conjecture unless $m_{A'}>0.3$ eV \cite{Reece:2018zvv}, and the viable mass range may shrink further with improved CMB bounds on the scale of inflation (see however Ref.~\cite{Craig:2018yld}). 

In light of the extensive experimental searches and limitations of these existing mechanisms, it is important to explore alternative ideas of how DPDM may be realized in a consistent cosmology. In this letter we study a novel possibility, in which energy stored in coherent oscillations of an axion-like field $\phi$ can be efficiently transferred to $A'$ via a so-called tachyonic instability. Explosive particle production via tachyonic instabilities was first studied in the context of preheating \cite{Traschen:1990sw,Kofman:1997yn} (for reviews, see Refs.~\cite{Allahverdi:2010xz,Amin:2014eta}) and has been discussed as a mechanism to produce vector fields in the context of generating primordial magnetic fields \cite{Anber:2006xt,Fujita:2015iga,Adshead:2016iae}. Axion-like fields are a well-motivated  ingredient in many theories beyond the SM and numerous string constructions.  While they may be dark matter themselves, their abundance could be easily suppressed in the presence of couplings to hidden sector fields (such as a dark photon) \cite{Agrawal:2017eqm,Kitajima:2017peg}. Here we show that even in this case, an axion-like field can be closely tied to the cosmological origin of dark matter in the form of a dark photon. In what follows, we specify the details of our DPDM production mechanism, and discuss possible subsequent thermal histories. A key ingredient is the depletion of the residual $\phi$ relic that remains after its partial conversion to $A'$. This is achieved via couplings to the visible sector. We consider two possibilities, where the $\phi$ relic either comes to dominate the energy density of the universe, or remains subdominant to SM radiation, before being thermalized into the SM bath and eventually depleted. In each case, we find DPDM production consistent with all cosmological constraints for a wide range of $\phi$ and $A^{\prime}$ masses. Our findings strengthen the case for DPDM searches.

\newpar
{\bf Tachyonic instability.}---%
%
We consider an axion-like particle $\phi$ (referred to as the ``axion'' hereafter) with a homogeneous initial value $\phi_i$ after inflation and reheating. It starts coherent oscillations when the Hubble rate is comparable to its mass, $3H\sim m_\phi$, around 
\begin{equation}
T_{\rm osc} \simeq 0.3 \sqrt{m_\phi \Mpl} \,,
\end{equation}
assuming an SM radiation-dominated epoch, where $\Mpl=2.4\times 10^{18}\,$GeV. Our DPDM production mechanism centers around a coupling of $\phi$  to a dark photon~$A'$,  
\begin{equation}
\label{eq:Lagrangian_dark}
\mathcal{L}_{\phi A' A'} = \frac{\alpha_D}{8\pi f_D} \phi\, F^{\prime}_{\mu\nu} \tilde{F}^{\prime\mu\nu} \,,
\end{equation}
with $F'$ being the dark photon field strength tensor and $\tilde{F}^{\prime\mu\nu} = \epsilon^{\alpha\beta\mu\nu} {F}^{\prime}_{\alpha\beta} /2$.   Here, we have introduced a fine structure constant $\alpha_D$ and axion decay constant $f_D$ for the dark sector. This interaction affects the dark photon equation of motion, written in Fourier space as \cite{Garretson:1992vt}
\begin{equation}
\label{eq:DP_EOM}
\frac{\partial^2 A_{\pm}' }{\partial\eta^2} + \left( m_{A'}^2 + k_{A'}^2 \pm \frac{\alpha_D k_{A'}}{2 \pi f_D} \frac{\partial \phi}{\partial\eta} \right) A_{\pm}' = 0 \,, 
\end{equation}
where $\eta$ is the conformal time, $m_{A'}$ and $k_{A'}$ are the mass and momentum of $A'$, and $\pm$ indicates helicity. We have assumed negligible $A'$ self-interactions; this implies important constraints \cite{Agrawal:2018vin} that we find may be always satisfied in the parameter space of interest.

Tachyonic instability refers to an efficient energy transfer from $\phi$ to $A'$, which occurs when $m_{A'}^2 + k_{A'}^2 \pm \frac{\alpha_D k_{A'}}{2 \pi f_D} \frac{\partial \phi}{\partial\eta} < 0$. This negative quantity can be regarded as a tachyonic effective mass for $A'$ and leads to an exponentially growing solution to Eq.~(\ref{eq:DP_EOM}). This tachyonic condition can be satisfied only after the axion starts rolling at $T_{\rm osc}$, i.e. $\partial \phi/\partial\eta \neq 0$. At a given time, one of the $A'$ helicities exhibits exponential growth, with a peak momentum $k_{A'}\sim \frac{\alpha_D}{4\pi f_D}\bigl|\frac{\partial\phi}{\partial\eta}\bigr|\gg m_{A'}$, obtained from minimizing the tachyonic effective mass. Efficient production via tachyonic instability requires $\frac{\alpha_D \phi_i }{2\pi f_D} \gtrsim \mathcal{O}(10)$ \cite{Agrawal:2017eqm}, which can be realized in a variety of setups accommodating $\phi_i \gg f_D$ \cite{Agrawal:2018mkd}. Tachyonic instability ceases when the produced $A'$ backreacts on the $\phi$ condensate to excite higher momentum modes so that $\phi$ is no longer described by a coherently oscillating field. Detailed numerical simulations are needed to accurately determine the final momentum distribution and yield. For the present study, we take $p_{A'}\sim p_\phi\sim m_\phi$, $n_{A'} = n_\phi \sim m_\phi \phi_i^2$ after backreaction sets in, consistent with the lattice results in Ref.~\cite{Kitajima:2017peg}. The uncertainty in these quantities should not affect our conclusions.

Shortly after the era of tachyonic instability, $\rho_\phi$ will scale as nonrelativistic matter and quickly dominate over $\rho_{A'}$. In the absence of additional couplings, $\phi$ would either decay to $A'$ at a later time or survive until today. In the first case, DPDM might be dominantly produced from $\phi$ decay but tends to be too hot. In the second case, dark matter would be dominantly axions if $m_\phi \gg m_{A'}$~\cite{Agrawal:2017eqm}. In principle, these two species can coexist as comparable components if $m_\phi \simeq m_{A'}$ \cite{Agrawal:2018vin}. This comes at the cost of fine-tuning because these masses with wildly different origins can span many decades and have no a priori reason to be close. Here we instead consider the possibility that the residual $\phi$ is depleted via couplings to the visible sector, so that dark matter is composed of just the $A'$ produced from the tachyonic instability. The remaining sections concentrate on understanding this $\phi$ depletion, which is largely independent of the $A'$ production mechanism discussed above.

\newpar
{\bf Depletion of the axion relic.}---%
%
We consider a setup where $\phi$ couples to the $U(1)_Y$ hypercharge gauge boson,
\begin{equation}
\label{eq:LphiBB}
\mathcal{L}_{\phi BB} = \frac{\alpha_Y}{8\pi f_B}\phi\, B^{\mu\nu}  \tilde{B}_{\mu\nu} \,,
\end{equation}
induced by Peccei-Quinn (PQ) fermions $\psi_B$ charged under $U(1)_Y$, where $\alpha_Y\simeq 10^{-2}$. We will treat $f_B$ and $f_D$ as independent parameters, keeping in mind that each can be separated from the PQ breaking scale $f_\text{PQ}$ \cite{Agrawal:2018mkd}. 

An example of how hierarchical decay constants could be realized is a clockwork theory \cite{Choi:2014rja,Choi:2015fiu, Kaplan:2015fuy,Long:2018nsl} where PQ fermions charged under different gauge groups reside on different sites. As a result, it is possible to have $f_\text{PQ} \lesssim f_{B,D} \ll f_a$ with either ordering of $f_B$ and $f_D$. Here $f_a$ is the inverse axion coupling to a postulated dark QCD sector on the last site, which sets the range of axion field excursions. Meanwhile, small explicit breaking of the clockwork symmetry can give additional contributions to $m_\phi$ without spoiling the clockwork mechanism, in which case $m_\phi$ is bounded only by $f_\text{PQ}$. In sum, we have
\begin{equation}
\label{eq:CWhierarchy}
m_\phi < f_\text{PQ} \lesssim f_{B,D} \ll f_a \sim \phi_i\,.
\end{equation}
Note that while our discussion assumes the hierarchy of Eq.~\eqref{eq:CWhierarchy}, we do not rely on the specific mechanism (e.g.\ clockwork) by which this hierarchy is realized.

We now discuss processes mediated by the $\phi BB$ coupling of Eq.~\eqref{eq:LphiBB} that may reduce the $\phi$ abundance by thermalizing $\phi$ into the SM bath. At high temperatures $T\gtrsim m_\phi/\alpha_Y^2$, the axion mass is smaller than the thermal width of $B$, and axion dissipation is resonantly enhanced via $\phi B\to B$, with a rate inversely proportional to the thermal width. Adapting the calculations in Refs.~\cite{Rychkov:2007uq,Salvio:2013iaa,Moroi:2014mqa} (see also Refs.~\cite{Yokoyama:2004pf,Mukaida:2012qn}) to the case of nonrelativistic $\phi$, we estimate the rate as
\begin{equation}
\label{eq:phiB2B}
\Gamma_{\phi B \to B} \simeq 10^{-3}\, \frac{m_\phi^2\, T}{f_B^2} \qquad (m_\phi/\alpha_Y^2 < T < m_{\psi_B})\,.
\end{equation}
Note the result is independent of $\alpha_Y$, since the $\alpha_Y^2$ from $\phi BB$ couplings is canceled by an $\alpha_Y^{-2}$ factor from the inverse thermal width. When $T$ falls below $m_\phi/\alpha_Y^2$, the resonance shuts off, and we instead need to consider $\phi B \to B^*\to f\bar f$. This process, however, has a rate that decreases more rapidly than the Hubble rate, and thus it cannot thermalize $\phi$ if $\phi$ was not thermalized already at higher temperatures. Finally, the decay $\phi\to BB$,  thermally blocked at high temperatures, opens up at $T\lesssim m_\phi$,
\begin{equation}
\label{eq:phi2BB}
\Gamma_{\phi\to BB} = \frac{\alpha_Y^2}{256\pi^3} \frac{m_\phi^3}{f_B^2} \qquad (T < m_\phi < m_{\psi_B})\,.
\end{equation}

This would be the whole story if $\psi_B$ were sufficiently massive to be decoupled at all times after $T_\text{osc}$. However, since $m_{\psi_B}\lesssim f_\text{PQ}$ for perturbative Yukawa couplings, decoupling $\psi_B$ would require $f_\text{PQ}$, and hence $f_B$ to be higher than $T_\text{osc}$, making thermalization of the $\phi$ relic very inefficient. Therefore, in the following we assume $m_{\psi_B}<T_\text{osc}$ so that lower $f_B$ values can be accommodated. We consider a minimal setup with just one species of $\psi_B$, vectorlike under $U(1)_Y$ with hypercharge $\pm 1$. A small amount of mixing with SM leptons, consistent with precision electroweak constraints, would allow $\psi_B$ to decay fast enough without having dangerous cosmological impact, see, e.g.\ Ref.~\cite{Kearney:2012zi} and references therein.

In this setup, $\psi_B$ itself can play a role in axion thermalization. After DPDM production at $T_\text{osc}$, there is a period when $T>m_{\psi_B},m_\phi$, and scattering processes involving the $\phi\psi_B\bar\psi_B$ coupling,
\begin{equation}
\label{eq:Lphipsipsi}
\mathcal{L}_{\phi \psi_B \bar\psi_B} = \frac{m_{\psi_B}}{f_B}\,\phi\, \bar\psi_B i\gamma^5 \psi_B \,,
\end{equation}
are dominant. We estimate the rate from
\begin{equation}
\label{eq:phiB2psipsi}
\Gamma_{\phi B \to \psi_B \bar\psi_B} \simeq 10^{-2} \,\alpha_Y \,\frac{m_{\psi_B}^2 T}{f_B^2} \qquad(T>m_{\psi_B},m_\phi)\,.
\end{equation}
Note that unlike the gauge boson case, axion dissipation via the $\phi\psi_B\bar\psi_B$ coupling cannot be resonantly enhanced because a fermion flips helicity after absorbing a (pseudo)scalar.
Subsequently, if $T$ falls below $m_{\psi_B}$ before reaching $m_\phi$, $\psi_B$ decouples and the previous discussion around Eqs.~\eqref{eq:phiB2B} and \eqref{eq:phi2BB} applies. Otherwise if $m_{\psi_B}<m_\phi/2$, Eq.~\eqref{eq:phiB2psipsi} holds until $T\sim m_\phi$, after which $\phi$ can decay to $\psi_B\bar\psi_B$ with a rate
\begin{equation}
\label{eq:phi2psipsi}
\Gamma_{\phi\to\psi_B\bar\psi_B} \simeq \frac{1}{8\pi} \frac{m_{\psi_B}^2 m_\phi}{f_B^2} \qquad(m_\phi> T,\, 2m_{\psi_B})\,.
\end{equation}

Eqs.~\eqref{eq:phiB2B}, \eqref{eq:phi2BB}, \eqref{eq:phiB2psipsi} and \eqref{eq:phi2psipsi} above, valid in different temperature regimes and for different mass orderings, allow us to relate $f_B$ to the thermalization temperature $T_\text{th}$, defined as the highest temperature at which $\Gamma(T)=H(T)$. Thermalization may happen either after or before the $\phi$ relic dominates the energy density of the universe, leading to different cosmological histories which we consider in the next two sections. Depending on the choice of parameters, further depletion may be needed after thermalization, which we will also discuss below.

\newpar
{\bf Matter-dominated universe.}---%
%
We first consider the possibility that $\phi$ comes to dominate the energy density of the universe and leads to a matter-dominated (MD) era starting at a temperature 
\begin{equation}
\label{eq:TM}
T_M \simeq 0.5 \,\frac{m_\phi^{1/2} \phi_i^2}{\Mpl^{3/2}} \,,
\end{equation}
before being thermalized at $T_\text{th}$ and reheating the universe. In this case, $T_\text{th}$ is fixed just by $m_\phi$ and $m_{A'}$, independent of the initial field value $\phi_i$. This is because the ratio $n_{A'} / n_\phi$, initially $\sim 1$ after DPDM production, is invariant until $T_\text{th}$, at which point $\rho_\phi=m_\phi n_\phi$ is converted into radiation energy density $\rho_\text{rad}$,
\begin{align}
Y_{A'} = \left. \frac{n_{A'}}{s} \right|_{T_\text{th}} \sim \left. \frac{n_\phi}{s} \right|_{T_\text{th}} = \left.\frac{\rho_\text{rad}}{m_\phi s}\right|_{T_\text{th}} = \frac{3}{4} \frac{T_\text{th}}{m_\phi}\,.
\end{align}
Using the observed dark matter abundance $\rho_{A'}/s = m_{A'} Y_{A'} \simeq 0.44$ eV, we thus obtain
\begin{equation}
\label{eq:TRMD}
T_\text{th} \simeq 0.6 \ \EV \frac{m_\phi}{m_{A'}}\equiv T_R \,,
\end{equation}
where we have introduced the notation $T_R$ to indicate that $T_\text{th}$ acts as a reheat temperature in the MD case.

If $m_{\psi_B}$ is given in addition to $m_\phi$ and $m_{A'}$, we would then be able to solve for $f_B$ from $\Gamma(T_R)=H(T_R)$. To ensure consistency, we need to check that for the $f_B$ value determined by $\Gamma(T_R)=H(T_R)$, $\Gamma(T)<H(T)$ at all temperatures $T>T_R$ where different processes may be relevant; otherwise the axion would have been thermalized earlier.  When performing this check, we maximize $H(T)$ by allowing matter domination to begin as early as possible, up to $T_\text{osc}$. This allows us to identify the maximal viable parameter space.

If a consistent solution for $f_B$ exists for given $(m_\phi$, $m_{A'}$, $m_{\psi_B})$, we then need to check if it satisfies the following constraints. First, since $m_\phi<f_\text{PQ}$, $m_{\psi_B}=y_{\psi_B}f_\text{PQ}/\sqrt{2}<(4\pi/\sqrt{2})f_\text{PQ}$ (assuming a perturbative coupling) while $f_\text{PQ}\lesssim f_B$, we need to impose at least
\begin{equation}
\label{eq:fBmin_mass}
f_B > m_\phi\,,\qquad
f_B > \frac{\sqrt{2}}{4\pi}m_{\psi_B}\,.
\end{equation}
Next, we require that $U(1)_\text{PQ}$ should not be restored by the thermal mass of the PQ breaking field, $S$, generated by $\psi_B$.   At minimum, this condition should be satisfied when $\phi$ starts to oscillate, i.e. $y_{\psi_B}^2 T_{\rm osc}^2/24+m_S^2 \sim m_{\psi_B}^2 T_{\rm osc}^2 /(12f_\text{PQ}^2) -f_\text{PQ}^2 <0$, where $m_S^2\sim -f_\text{PQ}^2$ for an ${\cal O}(1)$ quartic coupling. As a result,
\begin{equation}
\label{eq:fBmin_PQ}
f_B \gtrsim f_\text{PQ} \gtrsim 10^5 \ \GEV \left( \frac{m_\phi}{\GEV} \right)^{ \scalebox{1.01}{$\frac{1}{4}$} } \left( \frac{m_{\psi_B}}{100 \ \GEV} \right)^{ \scalebox{1.01}{$\frac{1}{2}$} } \,.
\end{equation}
To satisfy astrophysical constraints \cite{Ellis:1987pk,Raffelt:1987yt,Turner:1987by,Mayle:1987as,Raffelt:2006cw}, we require 
\begin{equation}
f_B > 10^7 \,\text{GeV} \quad \text{if}\;\; m_\phi < 10 \,\text{MeV}\,.
\end{equation}
Meanwhile, an upper bound on $f_B$ may apply if thermalization is not sufficient to deplete the $\phi$ relic, i.e.\ if the thermal abundance of $\phi$ at $T=T_R$ is higher than the dark matter abundance. This happens if $T_R>m_\phi$, in which case $\phi$ acquires a yield $1/g_*$ upon thermalization. We require that this thermal abundance should decay away fast enough so as not to dominate the energy density of the universe again, or to inject energy into the SM bath at late times, which may be subject to constraints from Big Bang Nucleosynthesis (BBN) or the CMB \cite{Kawasaki:1994sc,Fixsen:1996nj,Redondo:2008ec}. This requires $m_\phi > T_\text{BBN}$ (taken to be 3\,MeV), and $\Gamma_{\phi\to BB \text{ or } \psi_B\bar\psi_B} > H$ at $T=\max(m_\phi/g_*, T_{\rm BBN})\equiv T_{\text{d}}^{\min}$, which restricts $f_B$ to be
\begin{align}
\label{eq:fBmax_2ndTherm}
f_B \lesssim &
\left\{
\begin{array}{ll}
2 \times 10^8 \ \GEV  \sqrt{\frac{m_\phi}{\GEV}} \left( \frac{m_{\psi_B}}{T_{\text{d}}^{\min}}  \right) \quad 
&(m_{\psi_B}<\frac{m_\phi}{2})\,, \\
10^5 \ \GEV  \sqrt{\frac{m_\phi}{\GEV}} \left( \frac{m_\phi}{T_{\text{d}}^{\min}}  \right)\quad 
& (m_{\psi_B}>\frac{m_\phi}{2})\,,
\end{array}
\right. \nonumber\\
& \qquad\qquad\qquad\qquad\quad\text{applicable if  } T_R>m_\phi\,.
\end{align}

Further constraints on this scenario come from several requirements on $\phi_i$. First, $T_M>T_R$ implies
\begin{equation}
\label{eq:axion_dom}
\phi_i \gtrsim 2 \times 10^{15} \ \GEV \left( \frac{m_\phi}{\GEV} \right)^{ \scalebox{1.01}{$\frac{1}{4}$} }   \left( \frac{\mEV}{m_{A'}} \right)^{ \scalebox{1.01}{$\frac{1}{2}$} } \,.
\end{equation}
The second bound comes from the coldness of DPDM. The redshift of the DPDM momentum can be computed using the invariant ratios, $k_{A'}^3/n_\phi \sim m_\phi^2 / \phi_i^2$ until $T_R$, and $k_{A'}^3 /s$ after $T_R$. Requiring $k_{A'}<10^{-3}\,m_{A'}$ at $T =\,$eV \cite{Irsic:2017ixq,Lopez-Honorez:2017csg} and making use of Eq.~(\ref{eq:TRMD}), we obtain
\begin{equation}
\label{eq:DMwarmness}
\phi_i \gtrsim 4 \times 10^{10} \ \GEV \left( \frac{m_\phi}{\GEV} \right)   \left( \frac{\mEV}{m_{A'}} \right)^2 \,.
\end{equation}
Third, requiring that $\phi$ decay should not overproduce $A'$, $\Gamma_{\phi \rightarrow A'A'} \simeq \alpha_D^2 m_\phi^3/(256\pi^3 f_D^2) < H (T_R)$, we have
\begin{align}
\label{eq:DMfromTachIns}
\phi_i \gtrsim 10\cdot\frac{2\pi f_D}{\alpha_D} \gtrsim 10^6 \ \GEV \left( \frac{m_\phi}{\GEV} \right)^{ \scalebox{1.01}{$\frac{1}{2}$} }   \left( \frac{m_{A'}}{\mEV} \right) \,.
\end{align}
Finally, the isocurvature perturbation constraint from Planck \cite{Akrami:2018odb}, $\mathcal{P}_\text{iso}^{A'} \simeq\mathcal{P}_\text{iso}^\phi \simeq \left( \frac{H_I}{\pi \phi_i} \right)^2 \lesssim 8.7 \times 10^{-11}$, where $H_I$ is the Hubble rate during inflation, translates into
\begin{equation}
\label{eq:Iso}
\phi_i \gtrsim 3\times 10^4 \,H_I > 10^4 \,m_\phi\,.
\end{equation}
Here we have used $3H_I>m_\phi$, which is necessary for ensuring that $\phi$ starts oscillating only after inflation.

Lastly, we note there can be a contribution to DPDM from inflationary quantum fluctuations \cite{Graham:2015rva}, estimated as 
\begin{equation}
\label{eq:OmegaInf}
\frac{\Omega^{\rm inf}_{A'}}{\Omega_{\rm DM}} \simeq \frac{1}{D} \left( \frac{m_{A'}}{6 \times 10^{-6} \ \EV} \right)^{ \scalebox{1.01}{$\frac{1}{2}$} } \left( \frac{H_I}{10^{14} \ \GeV} \right)^2\,,
\end{equation}
where $D = \max(1, \sqrt{m_{A'}/H(T_R)})$ accounts for dilution from entropy production when $\phi$ is thermalized. Requiring Eq.~\eqref{eq:OmegaInf} to be less than 1, we obtain
\begin{equation}
\label{eq:DP_QaunFluc}
m_\phi < 3H_I \lesssim 8\times 10^{13}\,\text{GeV} \left(\frac{\text{meV}}{m_{A'}}\right)^{ \scalebox{1.01}{$\frac{1}{4}$} } D^{ \scalebox{1.01}{$\frac{1}{2}$} }\,.
\end{equation}

\begin{figure}[t]
\begin{center}
\includegraphics[width=\linewidth]{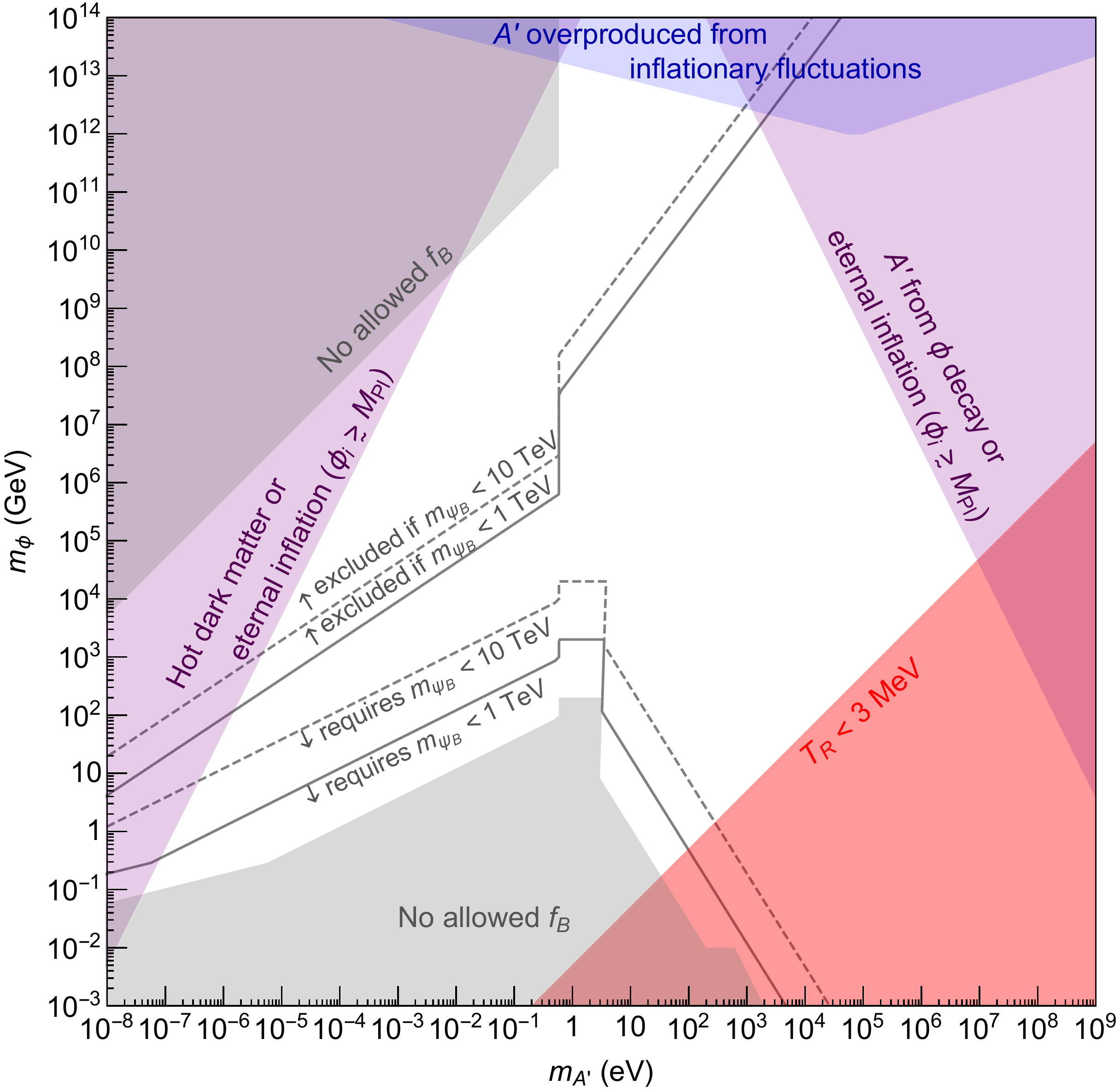}
\caption{
Parameter space of the dark photon and axion masses in the scenario where $\phi$ is thermalized after dominating the universe's energy density. 
}
\label{fig:MD_era}
\end{center}
\end{figure}

The aforementioned constraints are summarized in the $(m_{A'}, m_\phi)$ plane in Fig.~\ref{fig:MD_era}. The red region is excluded because $T_R<T_{\rm BBN}$, so the universe would be $\phi$-dominated at $T_{\rm BBN}$. In the purple region, the minimum $\phi_i$ consistent with Eqs.~(\ref{eq:axion_dom}-\ref{eq:Iso}) exceeds $\sqrt{2/3}\, \Mpl$, and the axion energy density would lead to eternal inflation. The blue region violates Eq.~\eqref{eq:DP_QaunFluc}, so DPDM would be overproduced from inflationary fluctuations.

We scan over $m_{\psi_B}$ (restricted to be above 100\,GeV to avoid possible tension with collider searches) to find the maximal parameter space allowed by Eqs.~(\ref{eq:fBmin_mass}-\ref{eq:fBmax_2ndTherm}), which excludes the gray regions. The irregular shape of these regions reflects the nature of this optimization procedure, namely adjusting $m_{\psi_B}$ to open the least constrained thermalization channel, among a set of options that differs from point to point in the $(m_{A'}, m_\phi)$ plane. For example, for $m_{A'}\gtrsim 0.6\,$eV, $T_R<m_\phi$ by Eq.~\eqref{eq:TRMD}, so we may choose between using Eq.~\eqref{eq:phi2BB} vs.\ Eq.~\eqref{eq:phi2psipsi} to thermalize $\phi$, while for $m_{A'}\lesssim 0.6\,$eV, the options are instead Eqs.~\eqref{eq:phiB2B} and \eqref{eq:phiB2psipsi}. In any case, we also need to ensure that $\Gamma(T)<H(T)$ for any $T>T_R$, which again requires considering different processes for different $m_{\psi_B}$. 

The availability of the remaining parameter space depends crucially on the PQ fermion mass. Interestingly, if we are to live in the region below the lower solid (dashed) curve (including the lowest $m_{A'}$ region our mechanism can accommodate), a concrete prediction would be the existence of a fermion with mass below $1(10)\,$TeV.  If a heavy lepton mixes with the SM as discussed above, it can decay to SM gauge bosons and fermions, a signature possibly accessible at the LHC or future colliders. Alternately, discovering a fermion with a sub-TeV mass would exclude all parameter space above the upper solid curve.

\newpar
{\bf Radiation-dominated universe.}---%
%
We now consider the second possibility, where the axion is thermalized before $T_M$, so the universe remains radiation dominated (RD). In this case, the DPDM relic abundance fixes $\phi_i$ rather than $T_\text{th}$. This is because in the absence of entropy production, the $A'$ yield is constant since $T_{\rm osc}$,
\begin{equation}
Y_{A'}  = \left. \frac{n_{A'}}{s} \right|_{T_{\rm osc}} \sim \left. \frac{n_\phi}{s} \right|_{T_{\rm osc}} \sim \, \frac{m_\phi \phi_i^2}{s(T_{\rm osc})}\,.
\end{equation}
The observed dark matter abundance then determines 
\begin{equation}
\label{eq:phiiRD}
\phi_i \simeq 2 \times 10^{15} \ \GEV \left( \frac{m_\phi}{\GeV} \right)^{ \scalebox{1.01}{$\frac{1}{4}$} }  \left( \frac{\mEV}{m_{A'}} \right)^{ \scalebox{1.01}{$\frac{1}{2}$} } \equiv \phi_0 \,,
\end{equation}
where we have introduced the notation $\phi_0$ to indicate that $\phi_i$ is fixed (given $m_\phi$, $m_{A'}$) in the RD case.

On the other hand, the axion thermalization temperature $T_\text{th}$ is not fixed, but subject to the following constraints. First, Eq.~\eqref{eq:phiiRD} combined with Eq.~(\ref{eq:TM}) fixes $T_M$, which sets a lower bound on $T_\text{th}$,
\begin{equation}
\label{eq:TMMD}
T_\text{th}> T_M \simeq 0.6 \ \EV \frac{m_\phi}{m_{A'}} \,. 
\end{equation}
Another lower bound comes from requiring that the $\phi$ relic should not decay to $A'$ before thermalization because this would give a potentially large contribution to hot dark matter. From $\Gamma_{\phi \rightarrow A'A'} < H (T_\text{th})$, we have
\begin{equation}
T_\text{th} \gtrsim 0.3\,\text{keV} \left( \frac{m_\phi}{\GeV} \right)^{ \scalebox{1.01}{$\frac{5}{4}$} }   \left( \frac{m_{A'}}{\mEV} \right)^{ \scalebox{1.01}{$\frac{1}{2}$} } \,,
\end{equation}
where we have used $\alpha_D/(2\pi f_D)\gtrsim 10/\phi_0$. Finally, we require $T_\text{th}>T_\text{BBN}$ to avoid potentially dangerous energy injection after BBN, and $T_\text{th}<T_\text{osc}$ so that $\phi$ is not thermalized when oscillations begin.

Once $T_\text{th}$ is chosen consistent with the aforementioned bounds, the procedure of determining $f_B$ and applying constraints on it is identical to the MD case (see Eqs.~(\ref{eq:fBmin_mass}-\ref{eq:fBmax_2ndTherm})), with an additional constraint from freeze-in (FI) overproduction of $A'$. Note that for $m_{A'} \gtrsim \mathcal{O}(100 \, \EV)$, the required dark matter yield is less than the thermal equilibrium value and hence any additional thermal production may result in an overabundance. In the current setup, a thermal abundance of $A'$ can freeze in via $\psi_B \bar\psi_B \to \phi^* \to A' A'$, which is most effective at the highest temperature $T_{\max}$. The FI abundance is computed to be (after using Eq.~\eqref{eq:phiiRD})
\begin{align}
\label{eq:FI}
\frac{\Omega^{\rm FI}_{A'}}{\Omega_{\rm DM}} \simeq  \left(\frac{\alpha_D\phi_0}{2\pi f_D}\right)^2 \left( \frac{m_{A'}}{\text{keV}} \right)^2 \left( \frac{m_{\psi_B}}{f_B} \right)^2  \left(\frac{T_{\max}}{T_\text{osc}}\right) \,.
\end{align}
At least, this ratio should be less than 1 for $\alpha_D\phi_0/(2\pi f_D)=10$ and $T_{\max}=T_\text{osc}$. Thus, 
\begin{equation}
f_B \gtrsim m_{\psi_B} \left(\frac{m_{A'}}{0.1\,\text{keV}}\right) \,.
\end{equation}
This constraint does not apply for the MD case as the FI abundance is sufficiently diluted by entropy production.

In addition, constraints from isocurvature perturbations and inflationary production, Eqs.~\eqref{eq:Iso} (with $\phi_i=\phi_0$ defined in Eq.~\eqref{eq:phiiRD}) and \eqref{eq:DP_QaunFluc} (with $D=1$) apply equally here. One last constraint is from the coldness of DPDM, $k_{A'}<10^{-3}\,m_{A'}$ at $T =\,$eV. The dark photon momentum is estimated using the invariant ratio $k_{A'}^3/s \sim m_\phi^3/s(T_{\rm osc})$, and the bound reads
\begin{equation}
m_\phi \lesssim 2 \times 10^6 \ \GEV  \left( \frac{m_{A'}}{\mEV} \right)^2 \,.
\end{equation}

\begin{figure}[t]
\begin{center}
\includegraphics[width=\linewidth]{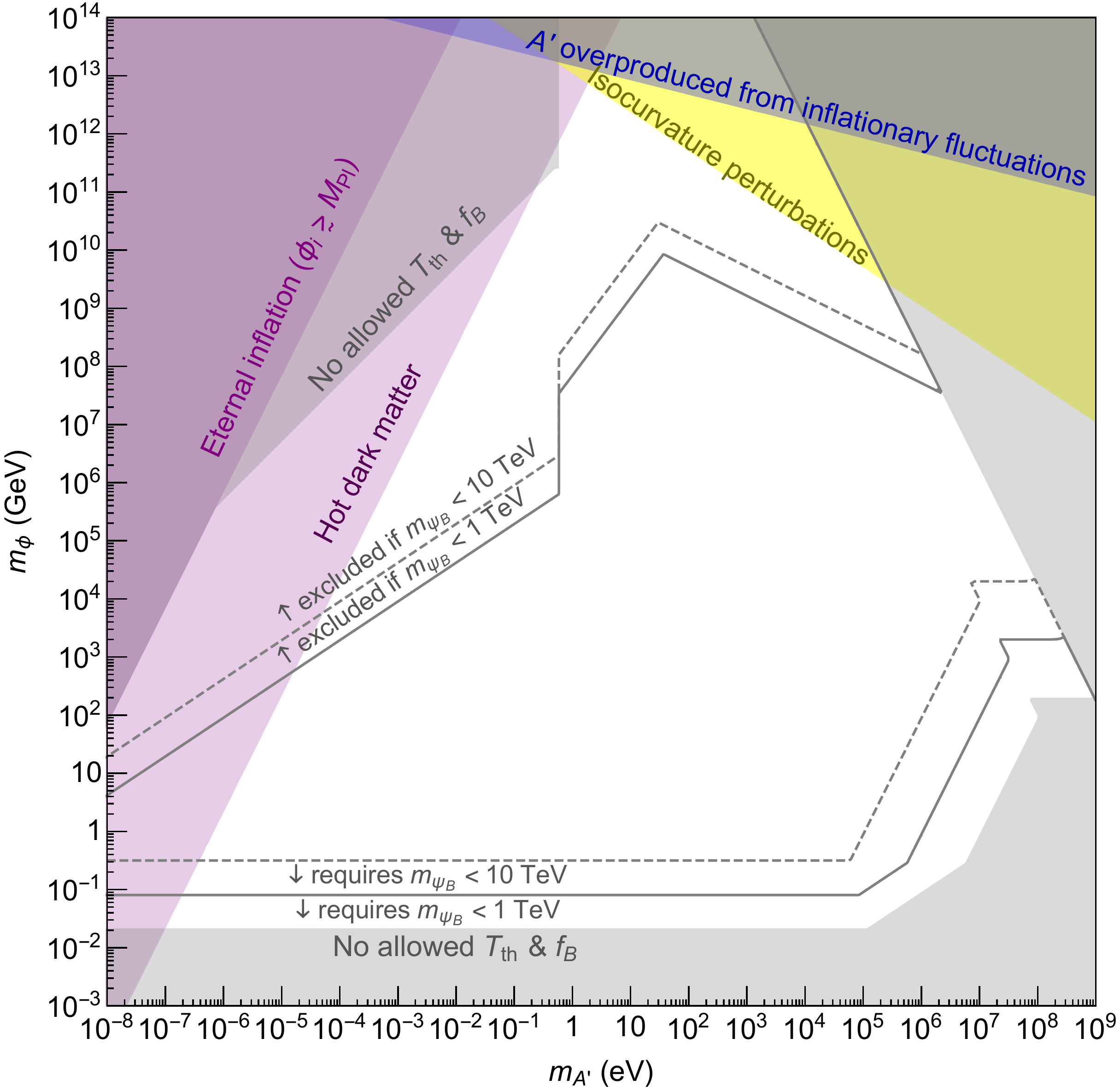}
\caption{
Similar to Fig.~\ref{fig:MD_era}, but for the scenario where $\phi$ is thermalized in a radiation-dominated universe.
}
\label{fig:RD_era}
\end{center}
\end{figure}

The result of all these constraints for this RD case is shown in Fig.~\ref{fig:RD_era}, where we scan over both $m_{\psi_B}$ and $T_\text{th}$ to find the maximal allowed parameter space. The additional freedom of adjusting $T_\text{th}$ opens up more parameter space compared to the MD case.

\newpar
 {\bf Discussion.}---%
%
We have shown that dark photon dark matter $A'$ produced by coherent oscillations of an axion-like field $\phi$ can have a viable cosmology for a broad swath of $A^{\prime}$ masses, ranging from a few $\times10^{-8}\,$eV up to $\sim$GeV. Two key ingredients are a tachyonic instability that leads to explosive production of $A'$ from $\phi$, and a mechanism to deplete the residual $\phi$ relic. For the latter, we have considered a minimal possibility of coupling $\phi$ to the visible sector as a proof of principle and discussed two possible thermalization histories, mapping out regions in the $(m_{A'}, m_\phi)$ plane consistent with all constraints (see Figs.~\ref{fig:MD_era} and \ref{fig:RD_era}). Our work shows that to explicitly realize the often-quoted ``nonthermal production'' of ultralight dark photon dark matter can be nontrivial, and motivates further investigation of the subject in light of the near-future experimental prospects of detecting such a dark matter candidate over a wide range of masses.

While our mechanism does not rely on interactions of $A'$ with the SM, a kinetic mixing $\epsilon$ with the photon is generically expected to exist, whose value is crucial for the potential detectability of DPDM. Constraints on $\epsilon$ specific to our mechanism include requiring that the $A'$-photon mixing should not disturb the momentum coherence necessary for the tachyonic instability, or thermalize the produced $A'$ at any time afterward. These constraints are however rather loose, because $\epsilon$ is suppressed at high temperatures by the large thermal mass of the SM photon. As a result, the only relevant constraints are those considered in Refs.~\cite{Arias:2012az, An:2013yua}, from a resonant conversion to SM photons. Experimental searches for DPDM will probe deeper into the $(m_{A'}, \epsilon)$ parameter space where our mechanism may be realized.

The most promising access to the PQ sector of our setup may be via a PQ fermion $\psi_B$ rather than the axion $\phi$ (which cannot be lighter than $\sim20\,$MeV and has a photon coupling that is already strongly constrained, see e.g.\ Eq.~\eqref{eq:fBmin_PQ}).  We have seen that the viable parameter space depends on $m_{\psi_B}$, because it affects the thermalization history of $\phi$. Interestingly, the lowest $m_{A'}$ region allowed by our mechanism (which goes beyond the mass range viable for inflationary production) is accompanied by $m_{\psi_B}\lesssim$ TeV. Discovery of such a PQ fermion at the LHC or a future collider, combined with positive results from light DPDM searches, may therefore hint toward an intertwined cosmological history of the PQ and dark photon sectors.


{\bf Acknowledgments.}---%
The authors thank K.~Harigaya, V.~Narayan, and B.~Safdi 
for useful discussions. The work of R.C.\ was supported in part by the DOE Early Career Grant No.\ DE-SC0019225. The work of A.P., Z.Z.\ and Y.Z.\ was supported in part by the DOE under Grant No.\ DE-SC0007859. Z.Z.\ was also supported by the Summer Leinweber Research Award, by NSF Grant No.\ PHY-1638509, and by DOE Contract No.\ DE-AC02-05CH11231. Z.Z.\ thanks the CERN theory group for hospitality during the completion of this work.

{\it Note added.}---%
During the preparation of this work, we became aware of Refs.~\cite{Agrawal:2018vin,Bastero-Gil:2018uel,Dror:2018pdh} which also consider new production mechanisms for light vector dark matter.

\end{document}